\documentclass[twocolumn, 11pt]{article}
\usepackage{amssymb,amsmath,amsfonts}
\usepackage{graphicx} 
\usepackage{bm} 
\usepackage{epsfig}

\title{Complex networks: new trends for the analysis of brain connectivity}
\author{MARIO CHAVEZ \\
\small{LENA-CNRS UPR-640. H\^{o}pital de la   Salp\^{e}tri\`{e}re.} \\
\small{47~Bd. de l'H\^{o}pital, 75651 Paris CEDEX 13, France} \\
\and
MIGUEL VALENCIA \\
\small{LENA-CNRS UPR-640. H\^{o}pital de la   Salp\^{e}tri\`{e}re.} \\
\small{47~Bd. de l'H\^{o}pital, 75651 Paris CEDEX 13, France} \\
\small{Department of Neurological Sciences, Center of Applied Medical Research,}\\
\small{University of Navarra, Avda Pio XII 31. 31008, Pamplona, Navarra} \\ 
\and
VITO LATORA \\
\small{Dipartimento di Fisica e Astronomia, Universit\`a di Catania and INFN,}\\
\small{Via S. Sofia, 64, 95123 Catania, Italy} \\ 
\small{Laboratorio sui Sistemi Complessi, Scuola Superiore di Catania, }\\
\small{Via San Nullo 5/i, 95123 Catania, Italy} \\ 
\and
JACQUES MARTINERIE \\
\small{LENA-CNRS UPR-640. H\^{o}pital de la   Salp\^{e}tri\`{e}re.} \\
\small{47~Bd. de l'H\^{o}pital, 75651 Paris CEDEX 13, France} \\}
\date{Last revised version: \today}

\begin{document}

\maketitle

\begin{abstract}
  Today, the human brain can be studied as a whole.
  Electroencephalography, magnetoencephalography, or functional
  magnetic resonance imaging (fMRI) techniques provide functional
  connectivity patterns between different brain areas, and during
  different pathological and cognitive neuro-dynamical states.  In
  this Tutorial we review novel complex networks approaches to unveil
  how brain networks can efficiently manage local processing and
  global integration for the transfer of information, while being at
  the same time capable of adapting to satisfy changing neural
  demands.
\end{abstract}

\section{Introduction}

In recent years, complex networks have provided an increasingly
challenging framework for the study of collective behaviors in complex
systems, based on the interplay between the wiring architecture and
the dynamical properties of the coupled units~\cite{netInNatureI,
  netInNatureII}. Many real networks were found to exhibit small-world
features. Small-world (SW) networks are characterized by having a
small average distance between any two nodes, as random graphs, and a
high clustering coefficient, as regular lattices ~\cite{watts98,
  tononi98, sporns00, efficiency}. Thus, a SW architecture is an
attractive model for brain connectivity because it leads distributed
neural assemblies to be integrated into a coherent process with an
optimized wiring cost~\cite{economic,complexNeuralNetsI, complexNeuralNetsII}.

Another property observed in many networks is the existence of a
modular organization in the wiring structure. Examples range from RNA
structures, to biological organisms and social groups. A module is
currently defined as a subset of units within a network such that
connections between them are denser than connections with the rest of
the network. It is generally acknowledged that modularity increases
robustness, flexibility and stability of biological
systems~\cite{Barabasi2004, Sole2008}. The widespread character of
modular architecture in real-world networks suggests that a network's
function is strongly ruled by the organization of their structural
subgroups.

Recent studies have attempted to characterize the functional
connectivity (patterns of statistical dependencies) observed between
brain activities recorded by electroencephalography (EEG),
magnetoencephalography (MEG), or functional magnetic resonance imaging
(fMRI) techniques~\cite{functionalNetsSFI, functionalNetsSFII,
  functionalNetsSFIII, functionalNetsSWI,
  functionalNetsSWII}. Surprisingly, functional connectivity patterns
obtained from MEG and EEG signals during different pathological and
cognitive neuro-dynamical states, were found to display SW
attributes~\cite{functionalNetsSWI, functionalNetsSWII}; whereas
functional patterns of fMRI often display a structure formed by highly
connected hubs, yielding an exponentially truncated power law in the
degree distribution~\cite{functionalNetsSFI, functionalNetsSFII,
  functionalNetsSFIII}. For a complete review of these issues, 
 reader can refer to the Refs.~\cite{complexBrainNets_I, complexBrainNets_II}.

In functional networks, two different nodes (representing two
electrodes, voxels or source regions) are supposed to be linked if
some defined statistical relation exceeds a threshold. Regardless of
the modality of recording activity (EEG, MEG or fMRI), topological
features of functional brain networks are currently defined over long
periods of time, neglecting possible instantaneous time-varying
properties of the topologies. Nevertheless, evidence suggests that the
emergence of a unified neural process is mediated by the continuous
formation and destruction of functional links over multiple time
scales~\cite{varela01, engel01, timScalesInNets}.

Empirical studies have lead to the hypothesis that transient
synchronization between distant and specific neural populations
underlies the integration of neural activities as unified and coherent
brain functions~\cite{engel01}. Specialized brain regions
would be largely distributed and linked to form a dynamical web-like
structure of the brain~\cite{varela01}.  Thus, brain  regions would be 
partitioned into a collection of modules, representing functional units, separable
from -but related to- other modules. Characterizing the dynamical
modular structure of the brain may be crucial to understand its
organization during different pathological or cognitive states. An
important question is whether the modular structure has a functional
role on brain processes such as the ongoing awareness of sensory
stimuli or perception.

To find the brain areas involved in a given cognitive task, clustering
is a classical approach that takes into account the properties of the
neurophysiological time series. Previous studies over the mammalian
and human brain networks have successfully used different  
methods to identify clusters of brain activities. Some classical
approaches, such as those based on principal components analysis (PCA)
and independent components analysis (ICA), make very strong
statistical assumptions (orthogonality and statistical independence of
the retrieved components, respectively) with no physiological
justification~\cite{Biswal1999, McKeown2003}.

In this Tutorial, we review an approach that allows to characterize the dynamic
evolution of functional brain networks~\cite{valencia2008, valencia2009}. We illustrate this approach on 
connectivity patterns extracted from MEG data recorded 
during a visual stimulus paradigm. Results reveal that the brain
connectivity patterns vary with time and frequency, while maintaining a
small-world structure. Further, we are able to reveal a non-random
modular organization of brain networks with a functional significance
of the retrieved modules.  This modular configuration might play a key
role in the integration of large scale brain activity, facilitating
the coordination of specialized brain systems during a cognitive brain
process.

\begin{figure*}[!htbp]
   \centering
   \resizebox{\columnwidth}{!}{\includegraphics{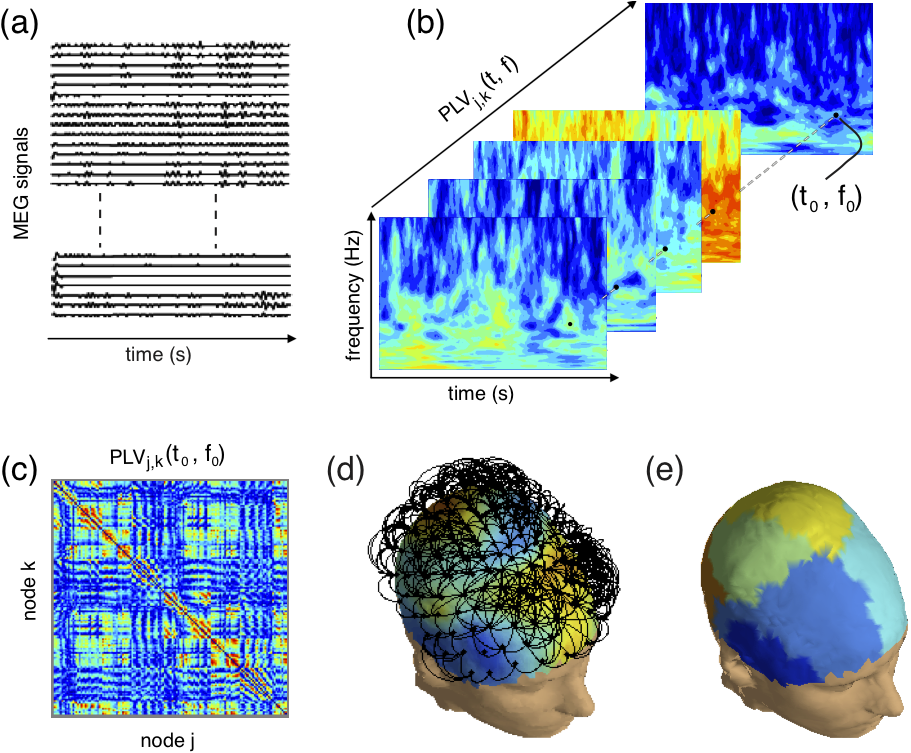}}
   \caption{General scheme for the extraction of the time-varying
     brain networks: (a) signals are decomposed into time-frequency
     components to compute (b) pair-wise relations; (c) functional
     connectivity matrices are extracted at each point of the
     time-frequency space, defining (d) the functional brain networks
     used to extract the topological attributes (color codes the nodes
     degree) and the (e) modular structure (brain sites belonging to
     each module are arbitrarily colored). See details in the text.}
    \label{FIGURE1}
 \end{figure*}

\section{Materials and methods} 

To illustrate our approach, we consider the brain responses recorded
during the visual presentation of non-familiar pictures. Although our
approach is applicable to any of the functional methods available
(EEG, fMRI, MEG), here we use the magnetoencephalography. This
modality of acquisition has the major feature that collective neural
behaviors, as synchronization of large and sparsely distributed
cortical assemblies, are reflected as interactions between MEG
signals~\cite{hamalainen93}.  We study the functional
connectivity patterns associated with dynamic brain processes elicited
by the repetitive application (trials) of a external visual
stimulus~\cite{ERPrefs}. For this experiment, a collection of $48$
simple structural images and scrambled images were randomly shown to
epileptic patients for a periode of
$150$~ms with an inter-stimulus interval of $2$~s. Patients were
required to respond by pressing a button each time an image was
perceived. The event-related brain responses were recorded (from two
patients) with a whole-head MEG system (151 sensors; VSM MedTech,
Coquitlam, BC, Canada) digitized at $1.25$ kHz with a bandpass of
$0-200$~Hz.

The basic steps of our approach are schematically illustrated in
Fig.~\ref{FIGURE1}. Each of the signals is decomposed into
time-frequency components, as shown in panel a).  The relations
between two signals $j$ and $k$ are firstly defined in time-frequency
space, as shown in panel b).  A statistical criterion is then used to
define a functional connectivity matrix for each time-frequency point,
panel c).  The details of the statistical criterion we adopted are
reported in Section \ref{efc}.  In panel d), topological metrics are
extracted from the connectivity patterns to obtain a time-frequency
characterization of brain networks. The metrics investigated are
analyzed in Section \ref{tvsbn}.  Finally, in panel e), at a given
frequency, or time instant of interest, the modular structure is
characterized as discussed in Sections \ref{nm} and \ref{cnp}. 
To evaluate the features of brain connectivity, the
obtained functional networks are compared with equivalent regular and
random networks.

\subsection{Estimation of functional connectivity} 
\label{efc}
A unified definition of brain connectivity is difficult from the fact that the recorded dynamics reflect the activities of neural networks at
different spatial and temporal resolutions. Three types of connectivity are currently considered: anatomical (description of the physical connections between two brain sites), functional (defined by a temporal correlation between distant neurophysiological events) and effective (causal influence that a neural system may exert over another). Here, we consider the functional links in brain signals defined by means of the phase-locking value (PLV) computed between all pairs of sensors~\cite{lachaux99}.  To compute the PLV values, we used a
complex Morlet's wavelet function defined as $w(t,f_0) =
A\exp(-t^{2}/2\sigma_{t}^{2})\times \exp(i2\pi f_0 t)$. Normalization
factor $A$ was set to $A=(\sigma_t \sqrt{\pi})^{-1/2}$. 
$\sigma_t = m/2\pi f_0$, $m$ is a constant that
defines the compromise between
time and frequency resolution, and
$f_0$ is the center frequency of the
wavelet. Hence, in time domain, its
real and imaginary parts are a
cosine and a sine, respectively, of
which the amplitude envelope is a
Gaussian with a standard deviation
of $\sigma_t$. In frequency domain, the
Morlet wavelet is also a Gaussian
with a standard deviation $\sigma_f$ given
$m=f_0/\sigma_f$. Here, $m$ was chosen to be 7.
By means of this complex wavelet
transform an instantaneous phase $\phi_{i}^{trial}(t,f)$ is obtained
for each frequency component of signals $i=1, \ldots, M$ at each repetition 
of the stimulus (trial). The PLV between any pair of signals $(i,k)$ is inversely
related to the variability of phase differences across trials:
\begin{multline*}\label{PLVequation}
    \text{PLV}_{i,k}(t,f)  = \\
    \phantom{=} \frac{1}{N_{trials}} \left| \sum_{trial=1}^{N_{trials}}
     \exp^{j (\phi_{i}^{trial}(t,f) - \phi_{k}^{trial}(t,f))} \right|,
\end{multline*}
where $N_{trials}$ is the total number of trials. If the phase difference varies little across trials, its distribution is concentrated around a preferred value and $\text{PLV} \sim 1$. In contrast, under the null hypothesis of a uniformity of phase distribution, PLV values are close to zero.

Finally, to assess whether two different sensors are \emph{functionally} connected, we calculated the significance probability of the PLV values by a Rayleigh test of uniformity of phase. According to this test, the significance of a PLV value determined from $N_{trials}$ can be calculated as $p=\exp(-N_{trials}\text{PLV}^2)$~\cite{fisherBOOK}. To correct for multiple testing, the False Discovery Rate (FDR) method was applied to each matrix of PLV values~\cite{FDR}. With this approach, the threshold of significance $PLV _{th}$ was set such that the expected fraction of false positives is restricted to $q \leq 0.05$.%

In the construction of the networks, a functional connection between two brain sites was assumed as an undirected and unweighted edge ($A_{ij} = 1$ if $PLV_{ij} > PLV_{th}$; and zero otherwise). Although topological features can also be straightforwardly generalized to weighted networks ~\cite{weightedLinks}, we obtained qualitative similar results (not reported here) for weighted networks with a functional connectivity strength between nodes given by $w_{ij} = PLV_{ij}$. More refined statistical tools can also be used to estimate 
time-varying and directed brain networks~\cite{fabrizio2008}.

\subsection{Time-varying structure of brain networks}
\label{tvsbn} 

A set of metrics can be used to characterize the topological
properties of the functional networks we have
constructed~\cite{netInNatureI, netInNatureII}. Here, we use three key
parameters: mean degree $\langle K \rangle$, clustering index $C$ and
global efficiency $E$. Briefly, the degree $k_i$ of node $i$ denotes
the number of functional links incident with the node and the mean
degree is obtained by averaging $k_i$ across all nodes of the
network. The clustering index quantifies the local density of
connections in a node's neighborhood. The clustering coefficient $c_i$
of a node $i$ is calculated as the number of links between the node's
neighbors divided by all their possible connections and $C$ is defined
as the average of $c_i$ taken over all nodes of the
network~\cite{watts98}. The global efficiency $E$ provides a measure
of the network's capability for information transfer between nodes and
is defined as the inverse of the harmonic mean of the shortest path
length $L_{ij}$ between each pair of nodes~\cite{efficiency}. The 
node-efficiency $E_i$ of the $i^{\text{th}}$ node is likewise defined
as the inverse of the harmonic mean of the minimum path length between
node $i$ and all other nodes in the network.

To asses the small-world behavior of functional networks, we perform a
benchmark comparison of the functional connectivity
patterns~\cite{watts98}. For this, the clustering and efficiency
coefficients of functional networks are compared with those obtained
from equivalent random and regular configurations. Regular networks
were obtained by rewiring the links of each node to its nearest (in
the sensors space) neighbors, yielding a nearest-neighbor connectivity
with the same degree distribution as the original network. To create
an ensemble of equivalent random networks we use the algorithm
described in Ref.~\cite{watts98}.  According to this procedure, each
edge of the original network is randomly rewired avoiding self and
duplicate connections. The obtained randomized networks preserve thus
the same mean degree as the original network whereas the rest of the
wiring structure is random.

\subsection{Network modularity} 
\label{nm}

Many real networks have a modular structure, i.e. their 
associated graphs are in general globally sparse but locally dense. In 
these networks, modules are defined as 
groups of vertices linked such that connections
between them are denser than connections with the rest of the
network. It is currently accepted that a partition $\mathcal{P} = \{
\mathcal{C}_{1},\ldots,\mathcal{C}_{M} \}$ represents a good division 
in modules if the portion of edges inside each module $\mathcal{C}_{i}$
(intra--modular edges) is high compared to the portion of edges
between them (inter--modular edges).  The modularity $Q(\mathcal{P})$,
for a given partition $\mathcal{P}$ of a network is formally defined
as~\cite{Newman2004}:
\begin{equation}
Q(\mathcal{P})=\sum_{s=1}^{M}{\left[{\frac{l_{s}}{L}-\left(\frac{k_{s}}{2L}\right)}^{2}\right]},
\end{equation} 
where $M$ is the number of modules, $L$ is the total number of connections in the network, $l_{s}$ is the number of connections between vertices in module $s$, and $k_{s}$ is the sum of the degrees of the vertices in module $s$.

To partition the functional networks in modules, we used a random walk-based algorithm~\cite{Pons2006}, because of its ability to manage very large networks, and its good performances in benchmark tests~\cite{Pons2006, Danon2005}.  Similar theoretical frameworks have been recently proposed for spectral coarse-graining~\cite{Allefeld2007, Gfeller2007}. The algorithm is based on the intuition that a random walker on a graph tends to remain into densely connected subsets corresponding to modules. Let $P_{ij}= \frac{A_{ij}}{k_{i}}$ to be the transition probability from node $i$ to node $j$, where $A_{ij}$ denotes the adjacency matrix and $k_{i}$ is the degree of the i$^{th}$ node. This defines the transition matrix $(P^t)_{ij}$ for a random walk process of length $t$ (denoted here $P^t_{ij}$ for simplicity). One can notice that, if two vertices $i$ and $j$ are in the same community, the probability $P^{t}_{ij}$ is high, and  $P^{t}_{ik}\simeq P^{t}_{jk} \forall k $.

The metric used to quantify the structural similarity between vertices
is given by
\begin{equation}
  \rho_{ij} = \sqrt{\sum_{l=1}^{N}\frac{(P^{t}_{il}-P^{t}_{jl})^{2}}{k_{l}}}
\end{equation}
This distance has several advantages: it quantifies the structural
similarity between vertices and it can be used in an efficient
clustering algorithm to maximize the network modularity $Q$. Further,
using matrix identities, the distance $\rho$ can be written as
$\rho^{2}_{ij}=\sum^{n}_{\alpha=2}{\lambda^{2t}_{\alpha}{(v_{\alpha}(i)-v_{\alpha}(j))}^{2}}$;
where $(\lambda_{\alpha})_{1 \leqslant \alpha \leqslant n}$ and
$(v_{\alpha})_{1 \leqslant \alpha \leqslant n}$ are the $n$
eigenvalues and right eigenvectors of the matrix $P$,
respectively~\cite{Pons2006}. This relates the random walk algorithm
to current methods using spectral properties of the
graphs~\cite{Newman2006, Gfeller2007}. The random-walk based approach,
however, needs not to explicitly compute the eigenvectors of the
matrix; a computation that rapidly becomes intractable when the size
of the graphs exceeds some thousands of vertices.

To find the modular structure, the algorithm starts with a partition in which each node in the network is the sole member of a module. Modules are then merged by an agglomerative approach based on a hierarchical clustering method~\cite{Ward1963}. The algorithm stops when all the nodes are grouped into a single component. At each step the algorithm evaluates the quality of partition $Q$. The partition that maximizes $Q$ is considered as the partition that better captures the modular structure of the network. In the calculation of $Q$, the algorithm excludes small isolated groups of connected vertices without any links to the main network. However, these isolated modules are considered here as part of the network for the calculation of the topological parameters.

\subsection{Comparison of network partitions} 
\label{cnp}
To evaluate the agreement between modules assignments at a given time
instant or frequency one can used the adjusted Rand index
$Ra$~\cite{Hubert1985}, which is a traditional criterion for
comparison of different results provided by classifiers and clustering
algorithms, including partitions with different numbers of classes or
clusters. For two partitions $P$ and $P'$, the original Rand index is
defined as~\cite{Rand1971} $R=\frac{a+d}{a+b+c+d}$; where $a$ is
number of pairs of data objects belonging to the same class in $P$ and
to the same class in $P'$, $b$ is number of pairs of data objects
belonging to the same class in $P$ and to different classes in $P'$,
$c$ is the number of pairs of data objects belonging to different
classes in $P$ and to the same class in $P'$, and $d$ is number of
pairs of data objects belonging to different classes in $P$ and to
different classes in $P'$. Thus the index $R$ has a straightforward
interpretation as a percentage of agreement between the two partitions
and it yields values between $0$ (if the two partitions are randomly
drawn) and $1$ (for identical partition structures).

The Rand index, however, has a bias if a partition is composed by many clusters, and it can take a non-null value for two completely random partitions. The index $R$ can be straightforwardly corrected for the expected value under the null hypothesis according to the following general scheme: $Ra=\frac{R-E\{R\}}{\max\{R\}-E\{R\}}$. Using the generalized hypergeometric distribution as the null hypothesis, the adjusted Rand index that corrects for the expected number of nodes pairs placed in the same module under two random partitions is given by~\cite{Hubert1985}
\begin{equation}
Ra=\frac{a-\frac{(a+c)(a+b)}{a+b+c+d}}{\frac{2a+b+c}{a+b+c+d}-\frac{(a+c)(a+b)}{a+b+c+d}}
\end{equation}
which has an expected value of zero under the null hypothesis, and it takes a maximum value of one for a perfect agreement of the two partitions. Thus, the adjusted Rand index is a statistics on the level of agreement or correlation between two partitions.

\section{Results} 
\subsection{Time-frequency dependence of brain networks} 
\begin{figure*}[!ht]
   \centering
   \resizebox{\textwidth}{!}{\includegraphics{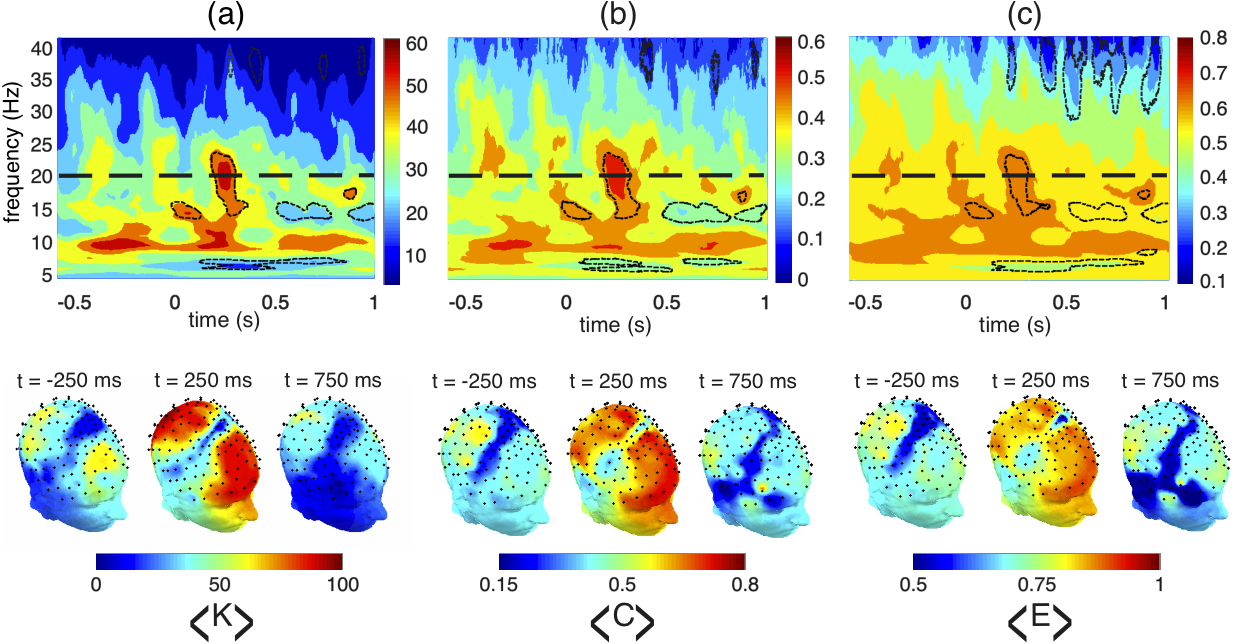}}
   \caption{Time-frequency maps of topological features extracted from brain
   networks associated to a visual stimulus presentation (arriving at $t=0$). (a) mean degree $\langle K \rangle$,
   (b) clustering index $C$ and (c) efficiency $E$. The reported values refer to the average over subjects.
   Dotted lines outline the regions revealing a significant change from the pre-stimulus region. 
   Lower row: topographic distribution of the local parameters for the 20 Hz activities (indicated by the thick dashed line) at three different time instants.}
    \label{FIGURE2}
 \end{figure*}

Fig.~\ref{FIGURE2} shows the topological attributes of functional networks elicited by the -unexpected- images. Pictures show the values of the mean degree, clustering index and efficiency of networks between, calculated at each point of the time-frequency space, $600$~ms before and $1$~s after the onset of the stimulus. 

The first crucial observation is that functional connectivity patterns are not time-invariant, but instead they exhibit a rich time-frequency structure during the neural processing. All the topological features (specially $\langle K \rangle$ and $C$) exhibit high values in a frequency band close to $10$~Hz, which is a spectral component mostly involved in the processing of visual information~\cite{ERPrefs}. Whereas the functional networks in the frequency range of $10-30$~Hz display large patterns of synchronization/desynchronization before the stimuli, a highly connected pattern is induced by the stimulus at about $250$~ms and between $15$ and $25$~Hz, suggesting a connectivity induced by the unexpected sensory stimuli. This is followed by weak connected structures at frequency bands close to $7$ and $15$~Hz arising during the post-stimulus activities and marking the transition between the moment of perception and the motor response of the subject. The topological features of these connectivity patterns were detected as statistically different from the pre-stimulus epoch by a $Z$-test corrected by a FDR at $q \leq 0.05$. Brain activities above $30$~Hz are characterized by a poor global connectivity. Local parameters, $k_i$, $c_i$ and $E_i$, for each sensor of the network are shown at three different time instants for a frequency of $20\ Hz$. During the processing of the stimulus, a time-space variability of connectivity is observed. Before the onset of the stimulus, the networks are characterized by a very sparse connectivity. Then, a clear clustered structure triggered by the stimulus appears at $t=250$~ms, defining two main regions (frontal and occipital) with a high density of connections. After the stimulus, the functional wiring displays again a sparse structure.
 
\subsection{Small-world behavior of brain networks} 
\begin{figure*}[!htbp]
   \centering
   \resizebox{\textwidth}{!}{\includegraphics{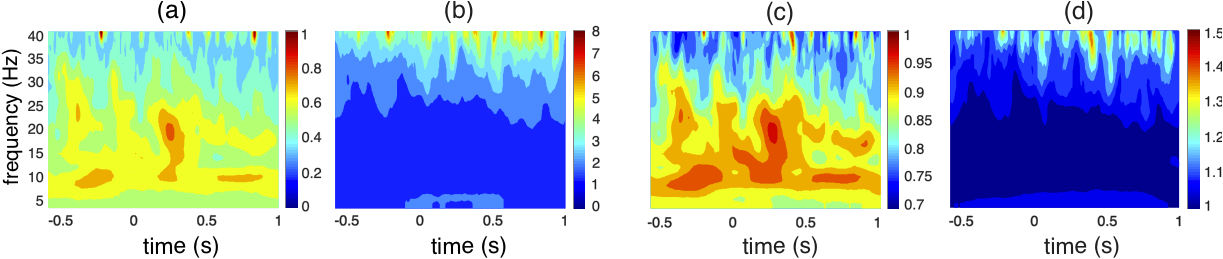}}
   \caption{Comparison of functional networks with random and regular
   configurations: time-frequency maps of (a) $C/\langle C_{lat}\rangle$,
   (b) $C/\langle C_{rnd}\rangle$, (c) $\langle E_{lat}\rangle/E$ and
   (d) $\langle E_{rnd}\rangle/E$. Results of equivalent random and regular
   networks refer to the average of $20$ realizations.}
    \label{FIGURE4}
 \end{figure*}

The comparison of the brain networks against random and regular configurations is shown Fig.~\ref{FIGURE4}. Typically, small-world
networks exhibit a $E_{sw}$ greater than regular lattices, but less than random wirings $E_{lat}< E_{sw} < E_{rnd}$; while for
the mean cluster index, $C_{rnd}< C_{sw} < C_{lat}$ is expected~\cite{watts98}. Results reveal that, despite the variability observed, functional networks display a topology different from regular and random networks. Namely, $\frac{C}{\langle C_{rnd} \rangle}>1$ and $\frac{C}{\langle C_{lat} \rangle} <1$, which indicates a SW structure ($\langle ...\rangle$ stays for an average over the ensembles of equivalent networks). Further, $\frac{\langle E_{lat} \rangle}{E}<1$ and $\frac{\langle E_{rnd} \rangle}{E}
> 1$, supporting the hypothesis of a SW connectivity.%

It is important to emphasize that, in contrast with previous studies which have focused on time-invariant networks~\cite{functionalNetsSFI, functionalNetsSFII, functionalNetsSFIII, functionalNetsSWI, functionalNetsSWII}, our approach reveals a \emph{dynamical} small-world connectivity at multiple time scales. This is a remarkable result, insofar as it suggests that the processing of a stimulus involves an optimized (in a SW sense) functional integration of distant brain regions by a \emph{dynamic} reconfiguration of links.

\subsection{Evolution of functional modules} 
\begin{figure*}[!htbp]
   \centering
   \resizebox{0.75\textwidth}{!}{\includegraphics{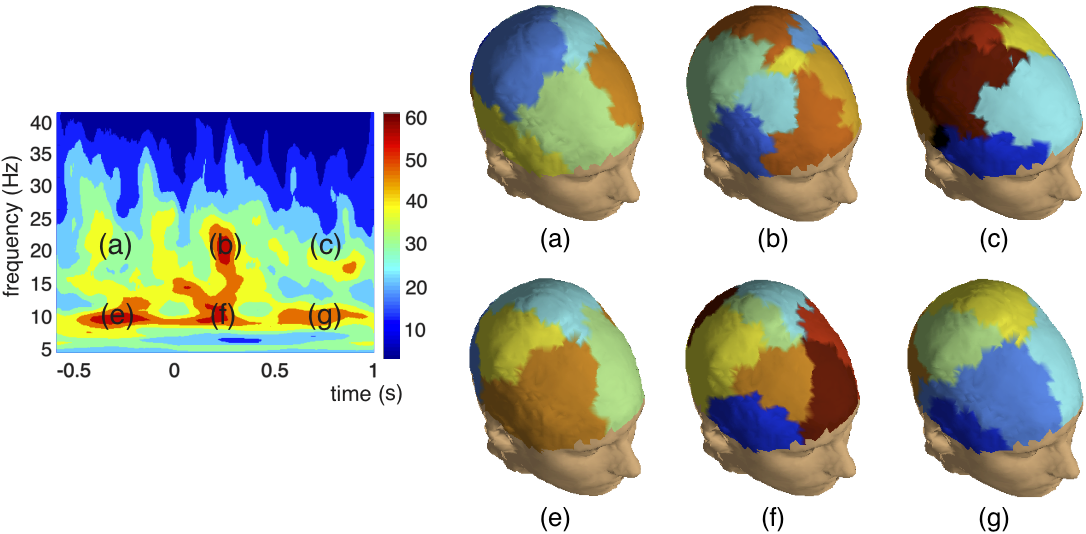}}
   \caption{Topographical distribution of the modules extracted from brain networks at different time instants and frequencies: (a) time instant t = -0.25 s, frequency f =20 Hz; (b) t = 0.25 s after the presentation of the stimulus at  f =20 Hz; (c) t =0.75 s, f =20 Hz; (e) t =-0.25 s, f =10 Hz; (f) t =0.25 s, f =10 Hz and (g) t =0.75 s, f =10 Hz. Brain sites belonging to each functional brain module were arbitrarily colored  (there is  no color correspondence between the modules of different networks). For the sake of clarity, isolated nodes were colored in black. (d) Time-frequency maps of mean degree is plotted to help network's localization in the time frequency space}
    \label{FIGURE5}
 \end{figure*}

A potential modularity of brain-webs is suggested by the fact that brain networks display a clustering index larger than that obtained from random configurations~\cite{Ravasz2002}.  Indeed, the presence of modules is actually confirmed by the high values of $Q$ obtained for brain networks extracted from brain activities at different time instants and frequencies. Fig.~\ref{FIGURE5} shows the 
spatial distribution of the modules for different networks with the following $Q$ values: (a) $Q=0.55$, (b) $Q=0.33$, (c) $Q=0.53$, (d) $Q=0.53$, (f) $Q=0.49$ and (g) $Q=0.53$. 

Results show that brain networks have a time-varying structure with a number of modules that changes with time and frequency. From plots one can observe that, despite the spatial variability observed at different time instants and frequencies, functional modules fit well some known brain regions including visual, somatosensory and auditory processing areas. Before the onset of the stimulus, however, networks are characterized by modular structures that define four main regions (anterior and posterior for both left and right hemisphere). We notice that the partitions of both networks present a relatively low agreement yielding a $Ra=0.39$. Then, the large connectivity triggered by the stimulus is also accompanied by an increase in the number of modules, yielding a more complex modular structure. The observed changes are directly related with the specific nature of the task:the detection and low-level processing of the stimulus involves the visual system, but further processing as the identification and perception of the picture requires
the mediation of regions as those located in frontal regions. Surprisingly antero-posterior relations elicit a large and unique module fitting fronto-occipital regions. Although a one-to-one assignment of anatomo-functional roles to each detected module is difficult to define, results reveal some other interesting modules, as the ones located over the motor cortex at $f = 18.5$~Hz. Then the post-stimulus activities recovers again a simpler spatial organization of modules.  It is worthy to notice that the pre- and post-stimulus networks have a very similar modular architecture only for the brain activities at the frequency band of $10$~Hz. This high agreement is  confirmed by a high value of the adjusted Rand index ($Ra=0.626$), compared with the values less than $0.38$ obtained for other frequencies.

These are remarkable results as they support the hypothesis that brain dynamics relies on different modular organizations to integrate distant specialized, but functionally related, brain regions.  Our findings suggest modularity as an organization basis leading distributed groups of specialized neural assemblies to be integrated into a coherent process during different cognitive or pathological states. A modular description of brain networks might provide, more in general, meaningful insights into the functional organization of brain activities during others neural functions, such as attention and consciousness.
 
\section{Conclusion}

In this Tutorial we have addressed a fundamental problem in brain
networks research: {\em whether and how brain behavior relies on the
  coordination of a dynamic mosaic of functional brain modules during
  cognitive states}.  We have proposed a method to study the
time-frequency dependencies of functional brain networks, thus
offering an \emph{instantaneous} description of the brain
architecture. Applied to a visual stimulus paradigm, the method
reveals that the functional brain connectivity evolves in a
small-world structure during the different episodes of the neural
processing. Furthermore, by using a random walk-based analysis, we have
identified a non-random modular structure in the functional brain
connectivity.

The present analysis was performed on MEG data in sensor space, which
contains some inherent spurious correlation between magnetic fields on
the surface of the brain. Although this caveat does not affect the
characterization of the global network topology, accurate inferences
about anatomical locations needs a source reconstruction of the
activity in the cortex. In this study, we have reduced the influence
of spurious correlations by simply excluding the nearest sensors from
the computation of PLV values.

Our approach may provide meaningful insights into how brain networks
can efficiently manage a local processing and a global integration for
the transfer of information, while being at the same time capable of
adapting to satisfy changing neural demands. Although the
neurophysiological mechanisms involved in the functional integration
of distant brain regions are still largely unknown, a dynamic SW
organization is a plausible solution to the apparently opposing needs
of local specificity of activity versus the constraints imposed by the
coordination of distributed brain areas. The modular structure
constitutes therefore an attractive model for the brain organization
as it supports the coexistence of a functional segregation of distant
specialized areas and their integration during brain
states~\cite{tononi98, sporns00}. We suggest that this network
description might provide new insights into the understanding of human
brain connectivity during pathological or cognitive states.

Applied to other multivariate data, our approach could provide new
insights into the structure of the time-varying connectivity at a
certain time~\cite{valencia2008}. A modular description of brain networks might provide, more
in general, meaningful insights into the functional organization of
brain activities recorded with others neuroimaging techniques 
(EEG, MEG or fMRI) during diverse cognitive or pathological 
states~\cite{valencia2009}.  In this study, the functional links have been defined in MEG signals by
means of the phase-locking value. We notice, however, that other
time-frequency methods (e.g. wavelet cross-spectra) can also be used
to detect and characterize a time-varying connectivity of spatially
extended, nonstationary systems (e.g. financial or epidemiological
networks). 

\section{Acknowledgments} 
The authors would like to thank S. Dupont, A. Ducorps and G. Yvert for clinical and technical support during data acquisition. This work was supported by the EU-GABA contract no. 043309 (NEST) and CIMA-UTE projects.

\end{document}